\documentclass[12pt]{article}
\usepackage[dvips]{graphicx}

\begin{document}
\title{\Large{\bf Yang-Lee Zeros of the Two- and Three-State Potts
Model Defined on $\phi^3 \, $ Feynman Diagrams }}
\bigskip
\author{{\bf Luiz C. de Albuquerque\thanks{e-mail:lclaudio@fatecsp.br}} \\
{\small\it Faculdade de Tecnologia de S\~ao Paulo - CEETEPS - UNESP }\\
{\small\it Pra\c{c}a Fernando Prestes, 30, 01124-060 S\~ao Paulo, SP,
  Brazil } \\
\\
{\bf D. Dalmazi\thanks{e-mail:dalmazi@feg.unesp.br }}\\
{\small\it UNESP - Campus de Guaratinguet\'a - DFQ }\\
{\small\it Av. Dr. Ariberto Pereira da Cunha, 333, 12516-410
  Guaratinguet\'a, SP, Brazil}}

\date{\today}
\maketitle
\begin{abstract}
  We present both analytic and numerical results on the position
  of the partition function zeros on the complex magnetic field plane of
  the $q=2$ (Ising) and $q=3$ states Potts model defined on $\phi^3 $
  Feynman diagrams (thin random graphs). Our analytic results are
  based on the ideas of destructive interference of coexisting phases
  and low temperature expansions. For the case of the
  Ising model  an argument based on a symmetry of the saddle point
  equations leads us to a nonperturbative proof that the Yang-Lee
  zeros are located on the unit circle, although no circle theorem is
  known in this case of random graphs. For the $q=3$ states Potts
  model our perturbative results indicate that the Yang-Lee zeros lie
  outside the unit circle. Both analytic results are confirmed by
  finite lattice numerical calculations.

\vfill
\noindent {\it PACS-No.}: 05.50.+q; 05.70.Fh; 64.60.Cn; 75.10.Hk

\end{abstract}
\clearpage

\newpage
\section{Introduction}

The Ising model is known to be exactly solvable in one dimension and
also in two dimensions in the absence of magnetic field. The magnetic 
field breaks the ${\cal Z}_2$ symmetry which complicates a possible
exact solution of the model. However, it is still possible to prove 
for arbitrary temperature an {\it exact} theorem ( see \cite{lee})
about the location of the zeros of the partition function on the complex
fugacity $u=e^{2H}$ plane, where $H \, $ stands for a complex magnetic 
field measured in units of $1/\beta = T\,$. Henceforth, these zeros 
will be called Yang-Lee zeros. The theorem establishes that the
Yang-Lee zeros are located on the unit circle $\vert u _k \vert =1 , 
\quad k=1,\cdots n\,$. Here $n$ is the number of sites of the lattice
whose details like coordination number and topology are not important 
for the proof. The symmetry under a change of sign of the magnetic
field $u\to 1/u $ is a key ingredient but it is not sufficient for the 
proof. The proof is a bit technical and strongly depends on the form
of the polynomials $P_n(u) $ in the partition function. A
simpler proof would certainly be welcome. The authors of \cite{biskup} 
have suggested an analytic way to find the location  of the partition 
function zeros. It is assumed that for a system defined in a periodic 
volume $\, V=L^d \, $ with $\, r$ different phases it is possible to 
write down the partition function of the system in terms
of some functions $f_l \,\,(l=1,\cdots , r\,)$ as follows:

\begin{equation}\label{1}
{\cal Z} \, = \, \sum_{l=1}^r q_l e^{-\beta V f_l} \, + \, {\cal
O}(e^{-L/L_0}e^{-\beta V f}).
\end{equation}
where $q_l$ is the degeneracy of the corresponding
phase $l$ and $f_l$ is interpreted as its metastable free energy. 
The quantity $L_0$ is of the order of the correlation length while 
$f={\rm min } {\Re e (f_k)}$. Following \cite{biskup}, we can use a 
destructive interference condition in order to
find the zeros of ${\cal Z}$ within ${\cal O}(e^{-L/L_0})$:

\begin{equation}
{\Re}e\,f_{l,L}^{{\rm eff}} \, = {\Re}e\,f_{m,L}^{{\rm eff}} \,
< \, \Re e\,f_{k,L}^{{\rm eff}} \quad  {\rm for}\; {\rm all } \quad
k\ne l,m \label{b1}
\end{equation}

\begin{equation}
\beta V \left( \Im m\,f^{{\rm eff}}_{l,L} - \Im m\,f^{{\rm eff}}_{m,L} 
\right) \, = \, \pi \quad {\rm mod} \quad 2\pi \label{b2}
\end{equation}
where $f_l^{\rm eff } = f_l - (\beta V)^{-1}\log q_l $. It
is assumed that the metastable free energies are functions of some 
complex parameter $\,z\, $ which is in our case the magnetic field. 
The continuity of the real part of the free energy across the line of 
zeros already appeared  in \cite{pearson}. It is
clear that equations (\ref{b1}) and (\ref{b2}) are useful whenever 
we have a closed expression for the free energies of the different 
phases of the system. In the case of the two dimensional Ising model 
in the presence of a magnetic field, although we do
not have exact expressions for the free energies we can use low 
temperature expansions (LTE's)  and the symmetry $\, H\to -H\, $ as  
in \cite{biskup} to furnish a simple proof of the Lee-Yang theorem 
valid in the thermodynamic limit and for low temperatures. In this
work we apply the same ideas for a particular case of
fluctuating random lattice. For such lattices no circle theorem 
has been proven. The spins degrees of freedom are placed on vertices 
of Feynman diagrams which are themselves dynamical degrees of freedom 
and need to be summed over in the partition function with appropriate 
combinatorial factors. In the case where the diagrams have
double lines and are generated by a random $\, N\times N \, $ matrix 
model (see \cite{kazakov,boulatov}) their sum  mimics, in the
continuum limit, the Ising model on random surfaces (two dimensional
gravity). Early \cite{staudacher,ambjorn} and recent
\cite{fatfisher} numerical results on graphs of planar topology 
indicate that the Yang-Lee zeros lie on the unit circle although 
no circle theorem exists in this case. Notice that the form of the 
polynomials used in the proof of the Lee-Yang theorem is
not preserved by their linear combinations. However, we believe that  
there might a generalization of the circle theorem for the new class 
of polynomials obtained by linear combinations with positive
coefficients of partition functions of the Ising type. In fact, 
further support to this conjecture has been given in \cite{aad} where
we studied the effect of the topology of the graphs on the position 
of the Yang-Lee zeros. We have noticed that changing from the planar 
to the torus topology the Yang-Lee zeros remained on the unit circle 
with a tiny change in their positions which is in agreement with the 
existence of an underlying robust theorem. Furthermore,
in order to take the continuum limit we take large matrices 
$\, N\to\infty \, $ which allows one to find (see \cite{kazakov}) 
with help of the orthogonal polynomial technique a closed expression 
for the free energy even in the presence of the magnetic field. Thus, 
an analytic study of the free energy singularities is possible and leads
to pure imaginary values for the magnetic field (see
\cite{staudacher}), i.e., $\vert u_k \vert =1$. Of course, this is 
only correct in the thermodynamic limit. On the other hand, the same 
issue of Yang-Lee zeros on a dynamical lattice can be studied in
a simplified model of thin graphs where matrices become numbers 
( $N=1$ ) and our final partition function becomes a linear 
combination of partition functions on single line Feynman diagrams. 
In this case no random surface interpretation exists and the
model exhibits a ferromagnetic phase transition with mean field 
exponents, see \cite{thinfisher} and references therein. 
Once again numerical results for finite number of vertices $\, n \,$  
indicate the unit circle as the locus of Yang-Lee zeros
\cite{aad}. However, since we do not have a random surface 
interpretation, the results of \cite{staudacher} do not go through 
the case of thin graphs. In particular, the location of the Yang-Lee 
zeros is not known even in the thermodynamic limit. In this
work  we analyze this question starting from a closed expression for 
the free energy of the model obtained in the thermodynamic limit 
by means of a saddle point approximation both for the Ising and the 
Potts model. Then, we use the destructive interference equations of 
\cite{biskup} to find out the location of the Yang-Lee
zeros. Unfortunately the presence of the magnetic field complicates 
the form of the resulting expression for the free energy but an 
argument based on  low temperature expansions and a symmetry 
($H\to -H$) between different saddle point solutions will
allow us to prove that the Yang-Lee zeros of the Ising model 
on thin graphs should be on the unit circle in the thermodynamic
limit. The proof is {\it exact} for the whole range of temperatures 
for which the LTE's converge. For the case of the
Potts model we have not been able to promote our low temperature 
results to an {\it exact } level but perturbatively we show that 
the Yang-Lee zeros should lie outside (but close to) the unit circle 
for low temperatures. The precise location of the zeros seem to change 
with the temperature in a complicated way. We also display numerical
results in agreement with our analytic findings.

\section{The Partition Function on Thin Graphs}

We can define the q-state Potts model as a generalization of the 
Ising model ($q=2 \, $) where on each site of the lattice there is 
a spin degree of freedom $\, \sigma \, $ which can take $\, q \, $ different
values $\, \sigma =1,2,\cdots , q $. By summing over all spin configurations
$\lbrace \sigma_i \rbrace $ on a given static lattice $\, G_n \, $
of $\, n \,$ vertices  we obtain the partition function 
$\, {\cal Z}_q(G_n) \,$. In the presence of a static magnetic field 
$ \, H \, $ in the $\sigma =1 $ direction we have,

\begin{equation}
{\cal Z}_q(G_n)\,=\, \sum_{\left\{\sigma_i\right\}}
 e^{\beta \sum_{<i,j>}
   \delta_{\sigma_i,\,\sigma_j}  \, + \,
   2H \sum_{i=1}^n \delta_{1,\,\sigma_i} } \, ,    \label{parcialz}
\end{equation}
where $\, \sum_{<i,j>}\, $ is a sum over all bonds ( propagators ) 
of $G_n$ including loop bonds which connect a site ( vertex ) to 
itself. Here we are interested in the case where the lattice is a 
dynamical degree of freedom and the final partition function is 
obtained by summing over all $G_n $ with $\, n \, $ vertices:
\begin{equation}
{\cal Z}_q^{(n)} \, = \, \sum_{k=1}^{K(n)} {\cal Z}_q (G_n^{(k)})
\label{finalz}
\end{equation}
In our case $K(n)$ will be the total number of Feynman graphs
$G_n^{(k)}$ with $\, n \, $ cubic vertices $\, \phi^3\, $ and no 
external legs. This clearly includes graphs of different topologies. 
Each ${\cal Z}_q (G_n^{(k)})\,$ as well  as  $\,{\cal Z}_q^{(n)}\,$ 
will be proportional to a polynomium $P_n(u)$ in the fugacity
$u=e^{2H}$. We take cubic vertices since those are the simplest 
ones after the trivial quadratic case $\, \phi^2 \, $  which 
corresponds to a $D=1$ lattice. The partition functions 
$\, {\cal Z}_q^{(n)} \, $ can be generated by integrating over $\,q\,$ 
zero dimensional fields $X_1,\cdots , X_q \, $ representing the 
$\, q \, $ different spin states $\, \sigma=1,2,\cdots , q \, $ 
respectively,

\begin{equation}\label{eq:s1}
{\cal Z}_q^{(2n)} \, = \frac{T_n}{2\pi\imath} \oint \frac{dg}{g^{2n+1}}
\left( \frac{\int {\cal D}\mu\, e^{-\frac12 \left\lbrack
 \sum_{i=1}^q X_i^2 \,-\, 2c \sum_{i>j} X_i X_j \,-\,
      \frac{2g}{3} (e^{2H} X_1^3 \,+\, \sum_{i=2}^q X_i^3)
\right\rbrack }}
{\int {\cal D}\mu \, e^{-\frac12 \left\lbrack
 \sum_{i=1}^q X_i^2 \,-\, 2c \sum_{i>j} X_i X_j \right\rbrack
}}\right)
\end{equation}
where ${\cal D}\mu = \prod_{i=1}^q  dX_i $ and $T_n$ is a numerical 
factor independent of the magnetic field and temperature which is 
necessary to get rid of the bad asymptotic behavior of the undecorated 
$\, \phi^3 \, $ graphs. The contour integral is
necessary to single out the term $\, g^{2n} \,$ containing
$\, 2n \, $ cubic vertices. Notice that $\,
{\cal Z}_q^{(n)}\, $ vanishes for odd number of vertices, therefore 
we write the total number of vertices  as $\, 2n \, $. 
The constant $\, c \, $ will be related to the temperature. 
To each cubic vertex ($X_1^3$) representing a spin in the direction of
the magnetic field there will be a $\, u=e^{2H} \, $ factor in 
agreement with the paramagnetic interaction term in (\ref{parcialz}). 
The quadratic terms in the argument of the exponentials in
eq. (\ref{eq:s1})  are responsible for the free propagators
$<X_i X_j>$ (or bonds $<\sigma_i \sigma_j>$ ) between the 
$\, q\, $ different vertices associates with the states $\sigma_i$. 
Taking ratios of the propagators and comparing with the ratios of 
the corresponding Boltzmann weights of (\ref{parcialz}) we identify
the parameter $c$ with a function of the temperature as follows. 
The action in the numerator of eq. (\ref{eq:s1}) can be written as
\begin{equation}\label{eq:s9} 
S_g\, =\, \frac 12\sum_{i,j=1}^q X_i
K_{ij} X_j - \frac{g}{3}\,( e^{2H}\, X_1^3 +
        \sum_{i=2}^q X_i^3)\,,
\end{equation}
where we have introduced the kinetic $q\times q$ matrix 
$K_{ij}=-c , {\rm if} \,i\ne j $ and $K_{ii}=1 $. 
We can check that $\det K = (1+c)^{q-1}(1-(q-1)c)$. The
propagators are given by the elements of the inverse matrix 
$K^{-1}_{ij}=(\det K)^{-1}c $ if $i\ne j$ and 
$K^{-1}_{ii}=(\det K)^{-1}\left[ 1-(q-2)c \right]$. Therefore, 
using the Boltzmann weights from (\ref{parcialz}) at vanishing 
magnetic field we have
\begin{equation}
\frac{\left\langle X_i X_j \right\rangle } {\left\langle X_i X_i
\right\rangle } \, = \, \frac{c}{\left[ 1-(q-2)c\right]} \, = \,
\frac{e^{E_{ij}(H=0)}}{e^{E_{ii}(H=0)}} \, = \, e^{-\beta} ,
\end{equation} 
and consequently

\begin{equation}
c\,=\, (e^{\beta}+q-2)^{-1} \, .\label{eq:s3}
\end{equation}
Therefore corresponding to $\, 0\le T\le \infty \, $ we have 
respectively the compact range $\, 0\le c\le 1/(q-1) \, $.

In the thermodynamic limit $\, n\to\infty \, $ one can evaluate the
free energy
\begin{equation}
f^{(2n)}_q \, = \, -\frac 1{2n}\, \log {\cal Z}_q^{(2n)} \label{f2n}
\end{equation}
by means of a saddle point approximation as follows. The fist step 
is to rescale the zero dimensional fields 
$\, X_i\to \frac 1g {\tilde X}_i \, $ such that $\, S_g(X_i,u,c)\to
(1/g^2)\, S_{g=1}({\tilde X}_i,u,c) \, $. Next, we decouple the 
contour integral over the coupling $\, g$  from the integral over 
the zero dimensional fields by a change of variables 
$\, g\to {\tilde g}\sqrt{S_{g=1}} \, $. Finally, we have
\begin{equation}\label{f}
f_q\, = \,
\lim_{n\to\infty}\left\lbrack -\frac1{2n}\, \log\, \int D\mu 
e^{-n\log S_{g=1}} \, + \, {\cal O}(\frac 1n )\right\rbrack
\end{equation}
Therefore the free energy per site is given in the thermodynamic 
limit by \cite{bachas}:

\begin{equation}
f_q\, =\, \frac 12\, \log\, {\tilde S}_{g=1} \label{ftilde}
\end{equation}
where $\, {\tilde S}_{g=1}\, $ corresponds to $\, S_{g=1} \, $ 
at a solution of the saddle point equations 
$\, \partial_{X_i}S_{g=1}=0 \, $. Henceforth it is always assumed
$g=1$ and we will treat the $q=2$ ( Ising ) and $q=3$ states Potts 
model separately.

\section{Ising Model}

Aligning the magnetic field in the direction of $\sigma_1$, which is 
represented by the field $X_1$, and  using the notation 
$\, X_1=x\, $ and $X_2=y\, $ we have respectively the saddle point 
equations $\, \partial_x S =0=\partial_y S \, $:

\begin{eqnarray}
x-cy\, &=& \, u x^2 \cr
y-cx\, &=& \, y^2  \label{4-1}
\end{eqnarray}

\noindent Using those equations we can write down the action at the 
saddle point in a quadratic form:
\begin{equation}
{\tilde S} \, = \, \frac 16\left\lbrack x^2 +y^2 -2cxy \right\rbrack
\label{4-2}
\end{equation}
From (\ref{4-1}) we have
\begin{equation}
x\, = \, (1/c)(y-y^2)\label{4-3}
\end{equation}
\begin{equation}
uy(1-y)^2 + c(y-1)+c^3 \, = \, 0 \label{4-4}
\end{equation}

\noindent In the absence of magnetic field ($u=1$) the solutions 
of (\ref{4-4}) simplify (see \cite{jpotts}) and we have two low 
temperature solutions below the critical value $c_{cr}=1/3\, $ 
and a high temperature solution valid for the entire range 
$0\le c\le 1\, $ :
\begin{eqnarray}
y_{LT\pm} \, &=& \, \frac 12 \left\lbrack 1+c \mp 
\sqrt{(1-3c)(1+c)}\right\rbrack \label{4-5}\\
y_{HT} \, &=& \,  1-c \label{4-6}
\end{eqnarray}

\noindent Defining the magnetization
\begin{equation}
m\, = \, \left\langle \frac{ux^3}{(ux^3+y^3)}\right\rangle \label{mq2}
\end{equation}
we can check that $y_{LT+}$ and $y_{LT-}$ correspond respectively 
to $m>1/2$ and $m<1/2$. Both solutions collapse at 
$m(x_{HT},y_{HT})=1/2$ for $c=1/3$. Substituting (\ref{4-5}), 
(\ref{4-6}) and (\ref{4-3}) back in (\ref{4-2}) we have:
\begin{eqnarray}
{\tilde S}_{LT} \, &=&\, \frac{(1+c)^2(1-2c)}{6} \quad \label{4-7}\\
{\tilde S}_{HT} \, &=& \, \frac{(1-c)^3}{3} \quad . \label{4-8}
\end{eqnarray}

\noindent Notice that both low temperature solutions  have 
furnished the same action ${\tilde S}_{LT}\, $.

The presence of a nonvanishing magnetic field ($\, u\ne 1 \, $) 
makes the solution of the cubic equation (\ref{4-4}) awkward and 
not very illuminating. We have solved (\ref{4-4}) instead as a 
Taylor expansion around $\, c=0 \, $ (LTE). We found three expansions
$y_{LT\pm}$ and $\, y_{HT} \, $ which reduce to (\ref{4-5}) and 
(\ref{4-6}) at $u=1$. They furnish the following actions :
\begin{eqnarray}
{\tilde S}_{LT+} \, &=& \, \frac 1{6u^2} - \frac{c^2}{2u^2} - 
\frac{c^3}{3u^3} + \frac{(u^2-1)c^4}{2u^4} + {\cal O}(c^5) 
\label{4-9}\\
&&\cr
{\tilde S}_{LT-} \, &=& \, \frac 1{6} - \frac{c^2}{2}-\frac{uc^3}{3} -
\frac{(u^2-1)c^4}{2} + {\cal O}(c^5) \label{4-10}\\
&&\cr
{\tilde S}_{HT} \, &=& \, \frac {1+u^2}{6u^2} - \frac cu +
\frac{(1+u^2)c^2}{2u^2} +\frac{(u^4-3u^2+1)c^3}{3u^3} \cr
&&\cr
&&\;\; +\, \frac{(u^2-1)^2(1+u^2)c^4}{2u^4} + {\cal O}(c^5)\label{4-11}
\end{eqnarray}

\noindent Clearly ${\tilde S}_{LT+} $ and ${\tilde S}_{LT-}$ 
collapse into (\ref{4-7}) at $u=1$. For $u>1 \, (H>0)  \,$ 
$f_{LT+}<f_{LT-}\, $ and the system is magnetized in the
direction of the magnetic field ($x$-direction ) while the situation 
is reversed for $u<1 \, (H<0)\, $ as expected. In order to locate 
the Yang-Lee zeros of ${\cal Z}_{q=2}^{(2n)}$ in the thermodynamic 
limit we first assume $u=\rho e^{\imath\theta}\,$ and then solve 
equations (2) and (3):

\begin{eqnarray}
& & \Re e\, (f_{LT+}-f_{LT-}) \,  =  \, 0 \label{4-12-a}\\
& & \Im m\, f_{LT-} \, = \, \Im m\, f_{LT+} + \frac{(2k-1)}
{\beta V \pi}\label{4-12-b}
\end{eqnarray}

\noindent There are two basic ingredients which will allow us to
locate {\it exactly} the Yang-Lee zeros for the whole convergence 
range of the LTE's without really worrying about the numerical details 
of the coefficients of the expansions (\ref{4-9}-\ref{4-11}).
First of all, notice that for $H\to -\infty $ (or $u\to 0\, $) the 
cubic equation (\ref{4-4}) has a unique solution 
$\, y(u\to 0)=1-c^2 \, $ and therefore there must be
at least one LTE which is well behaved for $\, u\to 0 \, $ and 
terminate at the term proportional to $c^2$. Of course, this can 
only be identified with  ${\tilde S}_{LT-}$ which implies that their 
coefficients can only involve positive powers of $\, u \,$, i. e. ,

\begin{equation} 
{\tilde S}_{LT-} (u,c)\, = \, \sum_{n=0}^{\infty } \, a_n(u) c^n\quad .
\label{4-13}
\end{equation}

\noindent Where each $a_n(u)$ is a polynomial in  $\, u \, $. 
The next important ingredient is a symmetry of the saddle point 
equations (\ref{4-1}) which relate solutions with $\, H>0 \, $ 
to solutions with $\, H<0 \, $. Namely, (\ref{4-1}) is invariant under

\begin{equation}
u \to 1/u
\end{equation}
\begin{equation}
x_i(1/u,c) \to u y_j(u,c) \label{4-14}
\end{equation}
\begin{equation} y_i(1/u,c) \to u
x_j (u,c)
\end{equation}

\noindent where $(x_i,y_i)$ and $(x_j,y_j)$ are, in principle, 
distinct solutions of (\ref{4-1}). Back in (\ref{4-2}) we are 
led to the relation:

\begin{equation}
{\tilde S}_i\, \Bigl(\frac{1}{u},c\Bigr)\, = \, u^2\,{\tilde S}_j
(u,c) \label{4-15}
\end{equation}

\noindent The labels $\, i,j \, $ indicate that the saddle point
solutions on both sides of (\ref{4-15}) do not need to be the same. 
From the first terms of the LTE's in (\ref{4-9})-(\ref{4-11}) and 
(\ref{4-15}) we have the identification below

\begin{eqnarray}
{\tilde S}_{LT+}(u,c) \, &=& \, \frac 1{u^2} \, 
{\tilde S}_{LT-}\left(\frac 1u,c\right) \label{4-16} \\
&&\cr
{\tilde S}_{HT}(u,c) \, &=& \, \frac 1{u^2} \, 
{\tilde S}_{HT}\left(\frac 1u,c\right) \label{4-17}
\end{eqnarray}

\noindent Putting back in the definition of the free energy and 
using (\ref{4-13}) we obtain

\begin{eqnarray}
& & 2(f_{LT-}-f_{LT+}) \, =\, \log u^2 \cr
&&\cr
&&\quad\quad +\;\left( \log\sum_{n=0}^{\infty} a_n(u)
c^n - \log\sum_{n=0}^{\infty} a_n\Bigl(\frac{1}{u}\Bigr) c^n\right)
\end{eqnarray}

\noindent Using $a_0=1/6$ and $u=\rho e^{\imath\theta } $, after 
expanding the logarithms we deduce:

\begin{equation}
2\Re e\, (f_{LT-}-f_{LT+}) \, = \, 2\log\rho + 
\sum_{n=1}^{\infty} c^n
\sum_{k=1}^{B(n)}A_{k,n}\cos (k\theta)\left(\rho^k - 
\frac1{\rho^k}\right)
\end{equation}

\noindent where $B(n)$ and $A_{k,n}$ are pure numerical factors. 
Thus, the condition $\,\Re e\, (f_{LT-}-f_{LT+})=0\, $ imply that 
the zeros must be on the unit circle $u_k=e^{\imath\theta_k}\,$ in 
the thermodynamic limit. The condition on the imaginary
parts of the free energies will locate the corresponding angles
$\theta_k $ on the circle as a function of the temperature. 
We emphasize that our proof of the circle theorem for thin $\phi^3$ 
graphs hold inside the convergence region of low temperature expansions.

\section{$q=3$ States Potts Model }

Using the notation $X_1=x,X_2=y,X_3=z$  and fixing the magnetic field 
once again along the $x$-direction we derive the saddle point
equations, $\, \partial_{X_i} S =0 \, $:

\begin{eqnarray}
x-c(y+z)\, &=& \, u x^2 \label{5-1} \\
y-c(x+z)\, &=& \, y^2 \label{5-2}\\
z-c(x+y)\, &=& \, z^2 \label{5-3}
\end{eqnarray}

\noindent From the difference of (\ref{5-2}) and (\ref{5-3}) we deduce 
that there are two groups of solutions, either 
$\, y=z\,$ or $\, y+z=1+c\, $ :

\vspace{1cm}

\noindent I)
\begin{eqnarray}
& & z\, = \, y \\
&&\cr
& & x\, = \, \frac yc (1-c-y) \label{5-4} \\
&&\cr
& & uy(y+c-1)^2 +c(y+c-1)+2c^3 \, = \, 0
\end{eqnarray}

\bigskip\bigskip

\noindent II)
\begin{eqnarray}
& & z\, = \, 1+c-y \\
&&\cr
& & x_{\pm} \, = \, \frac{1\pm \lbrack
1-4uc(1+c)\rbrack^{1/2}}{2u} \label{5-5} \\
&&\cr
& & y^2-(1+c)y+c(1+c+x)\, =\, 0
\end{eqnarray}

\noindent As in the Ising case it is instructive to first look at the 
solutions for vanishing magnetic field ($\, u=1 \, $). In this case 
we have from the first set of the solutions two low temperature and 
one high temperature solution :

\begin{eqnarray}
y_{LT\pm} \, &=& \, z_{LT\pm} \, = \, \frac{1\mp\lbrack
1-4c(1+c)\rbrack^{1/2}}{2} \cr
&&\cr
x_{LT\pm} \, &=& \, \frac{1+2c\pm \lbrack
1-4c(1+c)\rbrack^{1/2}}2 \label{5-6} \\
&&\cr x_{HT} \, &=& \, y_{HT} \, =  \, z_{HT} \, =\, 1- 2c
\end{eqnarray}

\noindent Defining the magnetization

\begin{equation}
m(q=3) \, =\, \left\langle \frac{ux^3}{ux^3+y^3+z^3}\right\rangle
\label{5-7}
\end{equation}

\noindent We identify the labels $LT+$ and $LT-$ with solutions such 
that $m > 1/3 $ and $m\le 1/3$ respectively. The solution $LT-$ 
meets the high temperature solution with $m=1/3$ at $c=1/5$. 
The second group of solutions (\ref{5-5}) does not furnish for
$u=1$ any new physical solution. Explicitly, for 
$\, 0\le c \le (\sqrt{2}-1)/2 \,$, we have from the second group 
two real solutions:

\begin{eqnarray}
x_1 \, &=& \, y_1 \, = \, \frac{1-\lbrack
  1-4c(1+c)\rbrack^{1/2}}2\label{5-8-a}\\
&&\cr
z_1 \, &=& \, \frac{1+\lbrack 1-4c(1+c)\rbrack^{1/2}}2 \label{5-8-b}
\end{eqnarray}
and

\begin{eqnarray}
x_2 \, &=& \, \frac{1+\lbrack
1-4c(1+c)\rbrack^{1/2}}2 \label{5-9-a} \\
&&\cr
y_2\, &=& \, \frac{1+c+\vert \lbrack
1-4c(1+c)\rbrack^{1/2}-c\vert }2 \label{5-9-b} \\
&&\cr
z_2\, &=& \, \frac{1+c-\vert
\lbrack 1-4c(1+c)\rbrack^{1/2}-c\vert }2 \label{5-9-c}
\end{eqnarray}

\noindent Notice that solution ($x_1,y_1,z_1$) corresponds to a 
permutation $(x,z)\to (z,x)$ of the solution $LT+$ given in
(\ref{5-6}) while the solution ($x_2,y_2,z_2$) coincides for 
$\, 0\le c \le 1/5 \, $ with the permutation $(x,z)\to ( z,x)$  of the
solution $LT-$. However, its derivative is discontinuous at $c=1/5$ 
and the solution deviates from $LT-$ for $1/5<c<(\sqrt{2}-1)/2\,$. 
Concluding, we can safely neglect both solutions of the second group.

Turning on the magnetic field ($\, u\ne 1 \, $), the solutions of 
the cubic equation in (\ref{5-4}) become cumbersome. Once again 
we make a low temperature expansion around $\, c=0 \, $ which we 
display below after the substitution in the action. From the first 
group of solutions ($\, y=z\,$) we have three possibilities: 

\begin{eqnarray}
{\tilde S}_0 \, &=& \, \frac{1+2u^2}{6u^2} - c \frac{2+u}{u^2} + 
\frac{c^2}{u^3}(1+2u+3u^2) \cr
&&\cr
&+& \frac{c^3}{3u^4} ( 8u^4 -13u^3-12u^2+2u+2) + {\cal O}(c^4)
\label{5-10}.
\end{eqnarray}

\begin{equation}
{\tilde S}_1 \, = \, \frac 1{6u^2} - \frac{c^2}{u^2}-
\frac{c^3(2+3u)}{3u^3} +{\cal O}(c^4) \label{5-11}
\end{equation}

\begin{equation}
{\tilde S}_2 \, = \, \frac 13 - c - c^2 + \frac{c^3(11-8u)}3 + 
{\cal O}(c^4)\label{5-12}
\end{equation}

\noindent Expanding the second group of solutions ($y+z=1+c $)  
we have two possibilities :

\begin{equation}
{\tilde S}_3 \, = \, \frac 16 - c^2 - \frac{c^3(u+4)}3 + {\cal O}(c^4)
\label{5-13}
\end{equation}

\begin{equation}
{\tilde S}_4 \, = \, \frac {1+u^2}{6u^2}-\frac cu  -
\frac{c^2}{u}-\frac{c^3(2+u)}{3} + {\cal O}(c^4) \label{5-14}
\end{equation}

\noindent Next we analyze  the physical content of all those
solutions. First, we notice that in the limit of vanishing magnetic 
field the only solution which correctly reproduces the LTE of 
$\,{\tilde S}_{HT}(u=1,c) \, $ is ${\tilde S}_0$ and therefore
we will identify it with the LTE of $\, {\tilde S}_{HT}(u,c)  \, $. 
However, the limit $\, u\to 1 \, $ can not be used to identify the 
remaining solutions since the solutions of the second group 
$\,{\tilde S}_3\, $  and $\, {\tilde S}_4 \, $ become identical
respectively to $\, {\tilde S}_1 \, $ and $\, {\tilde S}_2 \, $ 
for $\, u\to 1 \, $ which on their turn correspond to the LTE of the 
physical solutions ${\tilde S}_{LT+}$ and ${\tilde S}_{LT-}$ of the 
$\, H=0 \, $ case \footnote{Incidentally, we notice that
the discontinuous behavior of the solution (\ref{4-9})  only 
appears for $\,1/5<c<(\sqrt{2}-1)/2 \, $ which is outside the 
convergence region of the LTE of (\ref{4-9}). Remember that for 
$\, 0\le c \le 1/5 \, $ both ${\tilde S}_{LT-}$ and
(\ref{4-9}) are identical, which explains why ${\tilde S}_4 $ and
${\tilde S}_2$ become equal at $u=1$ and low temperatures.}.
In order to understand the meaning of ${\tilde S}_1-{\tilde S}_4$ 
we have to look at the limits $H\to +\infty $ ( $u\to\infty$)  
and $H\to -\infty $ ( $u\to 0 $ ). The only LTE solution well 
behaved for $u\to\infty $ is ${\tilde S}_1$  which
will be therefore identified with a strongly magnetized low temperature
solution, i.e., for this solution we expect $\lim_{c\to 0} m = 1 $.
For $u\to 0 \, $ both ${\tilde S}_2$ and ${\tilde S}_3$ are well 
behaved and good  candidates but ${\tilde S}_3$ possess the lowest 
free energy ( recall $f\sim \log\, S\, $ ) as one can check
numerically. Therefore ${\tilde S}_3$ will represent the weakly 
magnetized phase which satisfies $\lim_{c\to 0}m=0\, $. Thus, we have 
for the free energies of the physical low temperature solutions:

\begin{eqnarray}\label{5-15}
2(f_{LT-}-f_{LT+}) \, &=& \, \log\, {\tilde S}_3 - \log \,
{\tilde S}_1 \cr
&&\cr
&=& \, \log u^2 + \left(\frac 4u -2u -2 \right)c^3 \\
&&\cr
&+& 3c^4\left(\frac 2{u^2} -u^2 +\frac 4u - 2u
-3\right) \, + \, {\cal O}(c^5) \nonumber
\end{eqnarray}

\noindent We can find the position of the Yang-Lee zeros plugging 
$\, u=\rho e^{\imath\theta} \, $ in (\ref{5-15}) and solving 
the destructive interference equations (\ref{4-12-a}) and
(\ref{4-12-b}). First, if we look at the leading terms (
zero temperature ) we will see from the equations (\ref{4-12-a}) 
and (\ref{4-12-b}) that the Yang-Lee zeros, in the thermodynamic 
limit,  will be  located on the unit circle $\rho=1\, $ at $T=0$. 
Our finite size numerical calculations are in agreement
with those expectations (see Figure 1). Adding the next to leading 
terms of order $\,c^2 \, $ we will still have the zeros on the unit 
circle but if we further truncate the LTE´s at $\, c^3 \, $ level 
the zeros will slightly move out the unit circle. Indeed, from 
(\ref{4-12-a}) and (\ref{5-15}) we have at order $ \,c^3\, $ :
\begin{equation}
\cos
\theta \, = \, \frac{\rho\lbrack c^3 -\log\rho\rbrack}{c^3(2-\rho^2)}
\label{5-19}
\end{equation}
Imposing that $-1\le \cos\theta\le 1\,$ we find numerically for 
$\, 0\le c\le .5 \, $ that $\, 1\le \rho \le \rho^{*}\, $ 
where $\,\rho^{*} \, $ increases with the temperature. It is 
worth commenting that the r.h.s. of (\ref{5-19})  is a
monotonically decreasing function of $\rho$. Consequently, we have 
a closed curve outside the unit circle which is nonsymmetric 
across the imaginary axis with the farthest point from the origin 
being $(\rho=\rho^{*},\theta=\pi)$. In Figure 2 both our numerical 
and analytic results are overlapped. The finite size results seem to
tend to the (full line) analytic one in the thermodynamic limit. 
In Figure 2 we have used the truncation at order $\, c^4 \, $ 
instead of (\ref{5-19}), although they differ by a tiny amount, 
including the $c^4$ terms make the analytic results closer to
the numerical ones. The $c^4$ terms are quite complicated to be 
displayed  here explicitly. 

%%%%%%%%%%%%%%%%%%%%%%% FIGURE 1 %%%%%%%%%%%%%%%%%%%%%%%%%%%%%%%%%%%%%
\parbox{15cm}{
\begin{picture}(30,0)
\put(-15,37){{\footnotesize {\bf Fig.~1.} Yang-Lee zeros 
of ${\cal Z}_{q=3}^{(2n)}$ at zero temperature, $c=0$, for graphs 
with $2n=20$ ({$\diamond$}),  }}
\put(-15,25){{\footnotesize $2n=40$ ($\times$) and $2n=200$ ($\circ$) 
vertices. The closed curve is the unity circle.}}
\end{picture}
\includegraphics[width=10.2cm,height=14cm]{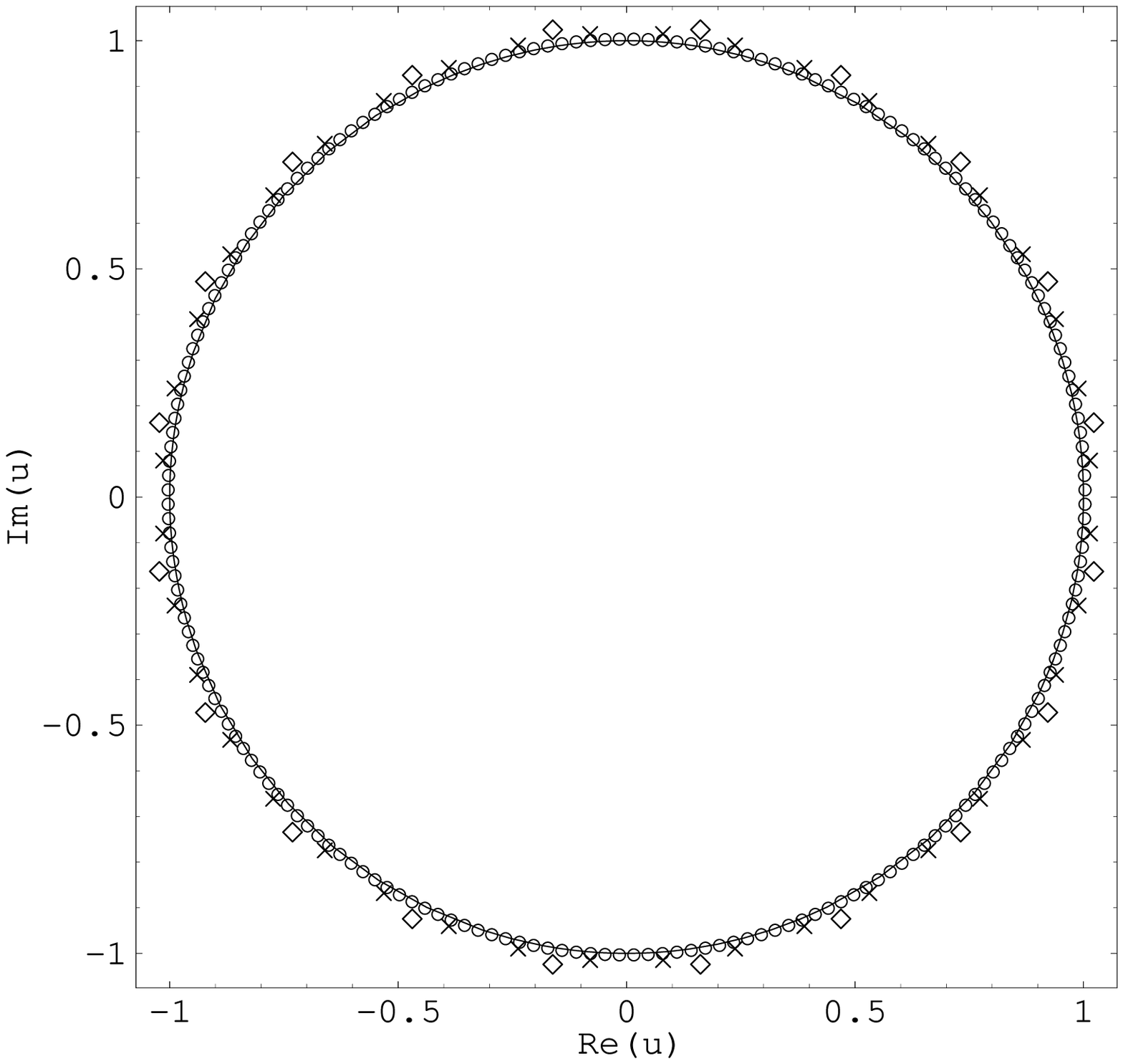}}
%%%%%%%%%%%%%%%%%%%%%%%%%%%%%%%%%%%%%%%%%%%%%%%%%%%%%%%%%%%%%%%%%%%%%

\section{$\, T \to \infty \, $}

The results of previous two sections hold for the range of 
temperatures for which the LTE's converge. In general, we have 
not been able to go beyond those temperatures analytically. 
One exception is the case $\, T\to\infty \, $ (or $c\to 1/(q-1)$) for
which a subtle decoupling of the zero dimensional fields take place 
for arbitrary $\,q\, $ similarly to what happens for the Ising model 
($q=2$), see \cite{aad}.  The key point is to take a particular 
scaling for the coupling $\, g\,$ appearing in the action
(\ref{eq:s9}) as \hfill\linebreak $c\to 1/(q-1)$. Namely, the matrix
$K_{ij}\,$ appearing in (\ref{eq:s9}) has $q-1$ degenerate eigenvalues 
$\lambda_j =1+c \,  (j=1,2, \cdots ,q-1)$, and the nondegenerate 
one $\lambda_q=1-(q-1)c$. In order to decouple the fields $X_j$ 
in the action it is natural to diagonalize the matrix
$K_{ij}$ through an orthogonal transformation\footnote{
Explicitly the orthogonal matrix reads
$A_{ij}=0 {~\rm if~} i\ge j+2 $ , $ A_{iq}=1/\sqrt{q}$
and $A_{ij}=- 1/\sqrt{j+j^2} {~\rm if~} i\le j\le q-1 \,{~\rm or~}
j/\sqrt{j+j^2} {~\rm if~} i=j+1\le q $.}
   $X_i\to A_{ij}X_j \,$
and rescale $X_i\to\frac{\tilde X_i}{\sqrt \lambda_i} $.
The Jacobian cancel out in the calculation of $\, {\cal Z}_q^{(2n)}\, $.
After those changes the action becomes (repeated indices are summed over)
\begin{equation}
  S_g\, =\, \sum_{i=1}^q \frac{{\tilde X}_i^2}2 \,-\, \frac{g}{3}
          \left\lbrack
     u\left(\frac{A_{1k}{\tilde X}_k}{\sqrt \lambda_k}\right)^3
\,+\,  \sum_{j=2}^q\left(\frac{A_{jk}{\tilde X}_k}
     {\sqrt \lambda_k}\right)^3 \right\rbrack \,.    \label{eq:s10}
\end{equation}

%%%%%%%%%%%%%%%%%%%%%%%%%% FIGURE 2 %%%%%%%%%%%%%%%%%%%%%%%%%%%%%%%%%%%%
\parbox{15cm}{
\begin{picture}(30,0)
\put(-15,37){{\footnotesize {\bf Fig.~2.} Yang-Lee zeros of 
${\cal Z}_{q=3}^{(2n)}$ at $c=0.2$, for graphs with $2n=20$ 
($\diamond$), $2n=40$ ($\times$) and $2n=200$ ($\circ$) }}
\put(-15,25){{\footnotesize vertices, compared to the saddle point 
result (solid line) and to the unity circle (dashed line).}}
\end{picture}
\includegraphics[width=10.2cm,height=14cm]{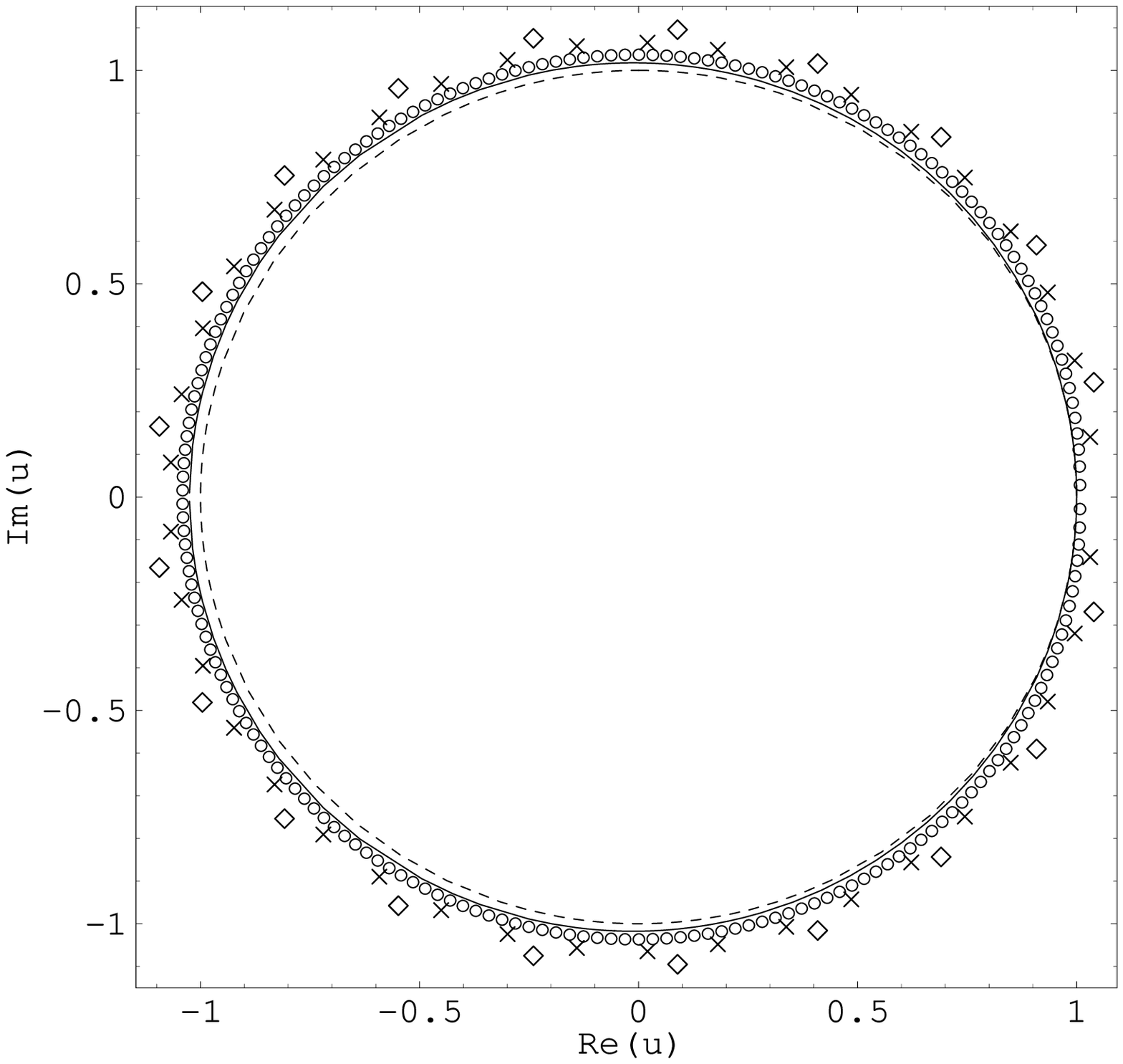}}
%%%%%%%%%%%%%%%%%%%%%%%%%%%%%%%%%%%%%%%%%%%%%%%%%%%%%%%%%%%%%%%%%%%%%

\noindent Apparently, we have just transferred the mixing terms to 
the higher powers of the potential, however in the limit 
$T\to\infty$ ($c\to 1/(q-1)$) we have $\, \lambda_q\to 0 \,$,
while all other eigenvalues remain finite. Thus, after the redefinition
$g\to {\overline g} \lambda_q^{3/2} $, we have in that
limit :
\begin{equation}
  S_g(T\to\infty)\, =\, \sum_{i=1}^{q-1} \frac{{\tilde X}_i^2}2 + 
\frac{{\tilde X}_q^2}2
-\frac{{\overline g}}{3}\Bigl( u\,A_{1q}^3 + \sum_{j=2}^q A_{jq}^3 \Bigr)
{\tilde X}_q^3 \, .           \label{eq:s11}
\end{equation}
Thus, we have decoupled the zero dimensional fields on
a specific point in the parameter space while keeping information
about nontrivial interacting terms. Clearly, this procedure is 
only useful if we do not care about the singular behavior of the 
original variable $X_q$. This is precisely the case of (\ref{eq:s1}).
Indeed, the Jacobian from $X_i$ to ${\tilde X}_i$ is singular in the
limit $T\to\infty$ but it cancels out because of the ratio in
eq. (\ref{eq:s1}). Since $A_{iq} = 1/\sqrt q$ we obtain
\begin{equation}
S(T\to\infty )\, =\,  \sum_{i=1}^q \frac{X_i^2}2 -
 \frac{{\overline g}}{3q^{3/2}}(y+q-1)X_q^3  \,. \label{eq:s12}
\end{equation}
Substituting this action in (\ref{eq:s1}), the integrations over
$X_1 ,\cdots ,X_{q-1}$ cancel out leaving us with an expression 
which is finite in the limit $\, c\to 1/(q-1) \, $. From the 
$\, {\overline g}^{2n} \, $ term we have :
\begin{equation}
{\cal Z}_q^{(2n)}(T\to\infty )  \, = \, {\tilde T}_n (u+q-1)^{2n}
\label{eq:s13}
\end{equation}
where $ {\tilde T}_n $ is a numerical factor independent of the 
magnetic field and temperature. Therefore we conclude that the 
Yang-Lee zeros of the $q$-state Potts model on thin graphs coalesce 
exactly at the point $u=1-q$ as $T\to \infty$. The same
result is valid on a static lattice \cite{creswick}. Again our 
numerical calculations confirm this analytic proof.

\section{Conclusion }

We have proved that in the thermodynamic limit the zeros of the 
partition function of the Ising model on thin graphs lie exactly 
on the unit circle in the complex fugacity plane.
Our proof is {\it exact} in the range of temperatures for which 
the low temperature expansions converge. For the case of the $q=3$ 
Potts model we do not have the $H\to -H$ symmetry anymore and
our results were fairly perturbative in the temperature. In that 
case the zeros lie on closed curves which depend on the temperature. 
Those curves lie outside the unit circle and tend to the circle as 
$T\to 0$. Our numerical results for small number of vertices seem to 
be in agreement with the analytic ones derived by means of the
destructive interference formulas of \cite{biskup}. The numerical 
results were very similar to the static lattice case treated in 
\cite{creswick} for both the Ising and $q=3$ Potts model. We should 
mention that part of the motivation for this work came from a recent 
progress (see \cite{thinfisher} and \cite{fatfisher}) on using the
destructive interference formulas of \cite{biskup} to find
analytically the exact curves formed by the Fisher zeros (complex 
temperatures) of the q-state Potts model on thin graphs. However, 
the results of \cite{thinfisher} and \cite{fatfisher} were
derived for $H=0$ and the presence of the magnetic field complicates 
the form of the solutions such that we had to appeal to low
temperature expansions.  As a further work one might look at Yang-Lee 
zeros for complex temperatures as in \cite{shrock} as well
as other types of vertices. Finally, we mention that, as in
\cite{aad}, we have also looked at connected partition functions 
obtained by first taking the logarithm of the ratio of integrals in 
(\ref{eq:s1}) and then making the contour integral. The results
for the position of the Yang-Lee zeros were qualitatively similar 
to the ones reported here. We hope to return in the future to the 
open problem of proving the circle theorem for the Ising model on 
random lattices with finite number of vertices.

\section{Acknowledgements}

The authors would like to thank Nelson A. Alves for useful discussions
and collaboration in early stages of  this work
and also Prof. D.J. Johnston for some useful e-mail exchange.
L.C.A. would like to thank the Mathematical Physics Department
of USP at S\~ao Paulo for their kind hospitality.
This work was partially supported by FAPESP, grants 00/03277-3
(L.C.A.) and 00/12661-1 (D.D.),  and CNPq (D.D.).

\vfill
%%%%%%%%%%%%%%%%%%%%%%%%%%%%%%%%%%%%%%%%%%%%%%%%%%%%%%%%%%%%%%%%%%%%%%%%%%%%%%%%%%%%%
\begin{center}
{\bf Figure captions}
\bigskip

\begin{itemize}

\item {\bf Fig. 1.} Yang-Lee zeros of ${\cal Z}_{q=3}^{(2n)}$
at zero temperature, $c=0$, for graphs with $2n=20$ ($\diamond$),
$2n=40$ ($\times$) and $2n=200$ ($\circ$) vertices.
The closed curve is the unity circle.

\bigskip

\item {\bf Fig. 2.} Yang-Lee zeros of ${\cal Z}_{q=3}^{(2n)}$
at $c=0.2$, for graphs with $2n=20$ ($\diamond$),
$2n=40$ ($\times$) and $2n=200$ ($\circ$) vertices, compared to the
saddle point result (solid line) and to the unity circle (dashed line).

\end{itemize}

\end{center}

%%%%%%%%%%%%%%%%%%%%%%%%%%%%%%%%%%%%%%%%%%%%%%%%%%%%%%%%%%%%%%%%%

\end{document}